\documentclass[
 reprint,
 amsmath,amssymb,
 aps,prl
]{revtex4-2}
\usepackage[usenames,dvipsnames]{xcolor}
\usepackage{graphicx}
\usepackage{svg}
\usepackage[hidelinks]{hyperref}
\usepackage{dcolumn}
\usepackage{bm}
\usepackage{hyperref}
\usepackage{physics, amsmath, bm}
\usepackage{tikz}
\usepackage{lipsum, xcolor}
\usepackage{subcaption}
\usepackage{ragged2e}

\begin{document}

\preprint{APS/123-QED}

\title{Emergent Long-Range Entanglement in the XXZ \\ Quantum Spin Chain with Competing Orders}

\author{Anthony Balchin\(^1\), Eytan Grosfeld\(^2\), Eran Ginossar\(^1\)}
\affiliation{\(^1\)\!School of Mathematics and Physics, University of Surrey, Guildford, GU2 7XH, UK. \\ \(^2\)\!Department of Physics, Ben-Gurion University of the Negev, Be'er Sheva 84105, Israel}

\date{\today}

\begin{abstract}

Competition between short-range and long-range interactions is a largely unexplored route to one-dimensional quantum matter. We introduce a spin-1/2 XXZ chain with ferromagnetic nearest-neighbour and antiferromagnetic power-law couplings. Using DMRG and spin-wave theory, we show this competition stabilises a gapless phase with central charge \(c=0.997(17)\), a twofold-degenerate ground state and entanglement spectrum, absent in nearest-neighbour models, and is long-range entangled as diagnosed by finite quantum conditional mutual information. The model also generates a unique gapped quantum phase and gapless antiphase with non-decaying correlations.  

\end{abstract}

\maketitle

\emph{Introduction}.--- Interactions of opposite sign, known as competing orders \cite{Defenu2023, Tahriri2025}, offer a  route to rich non-trivial physics by generating frustration \cite{Torre2025}. Even with short-range (SR) couplings, this competition can generate long-range (LR) entanglement \cite{Venuti2006} and multi-partite entanglement \cite{Jindal2014}. In addition, they can stabilise exotic phases distinguished by co-existing broken symmetries and non-local order such as valence-bond, chiral, and hidden topologically ordered phases \cite{Tahriri2025, agrapidis2019, Hikihara2008, Furukawa2012, Cea2024, Fumani2021, Ueda2014b, Wardyn2023}. 

However, competing orders have been studied separately for SR and LR, but rarely together \cite{callison2017, li2021, Sandvick2010}. Extensions to next-nearest neighbour interactions with negative-sign ferromagnetic (FM) interactions and positive-sign antiferromagnetic (AFM) interactions give rise to topological order \cite{agrapidis2019, Hikihara2008, Furukawa2012}. Such models with competing extended exchange interactions are also relevant to cuprates \cite{Moure2018, Laflorencie2005, Sun2017, Wang2021, Tang2024, Liu2026}. Broadly, physical systems realising LR interactions are experimentally realisable \cite{Defenu2023} with Rydberg atoms \cite{Saffman2010, Secker2016, Londono2025}, trapped ions \cite{Deng2005, Blatt2012, Gambetta2020}, and ultracold gases \cite{Bloch2008, Karman2018, Li2021UCG, Xie2022}. 

Competing SR-LR interactions therefore combines two sources of non-trivial physics: frustration from competing orders, and relaxed 1D constraints from the power-law decay. To study their interplay, we must consider relative oppositely signed SR as well as frustration-inducing LR anisotropic Heisenberg interactions Eq. \eqref{eq:hamG}. We find a rich phase diagram with several distinct phases in the LR dominated regime. They are characterised by relatively high level of correlation and signatures of LR entanglement, in contrast to earlier findings on related models with LR interactions with opposite overall signs \cite{maghrebi2017}.

\emph{The model}.--- We consider the following spin-1/2 LR-XXZ chain with open boundaries, \begin{eqnarray}\label{eq:hamG}
    \hat{H} = \sum_{i<j}^L\frac{G(i,j)}{\abs{i-j}^\alpha} \left[ S_i^x S_{j}^x + S_i^y S_{j}^y +\beta S_i^z S_{j}^z \right],
\end{eqnarray} 
where \(S_i^{x,y,z} = \frac{1}{2}\sigma_i^{x,y,z}\), \(\beta \in \mathbb{R}\), and the model has \(U(1)\times\mathbb{Z}_2\) symmetry, enhanced to \(SU(2)\) at \(\beta=1\). The range function is \(G(i,j)=-1\) for \(\abs{i-j}=1\) and \(G(i,j)=1\) otherwise, so SR nearest-neighbour (NN) pairs have negative-sign and the LR power-law couplings are positive-sign. We compare against the well-studied SR regime \cite{giamarchi} and non-competing LR chain \cite{maghrebi2017, Frerot2017, Ren2020, schneider2022, Moure2018}. 

\begin{figure}[h!]
\includegraphics[width=0.48\textwidth]{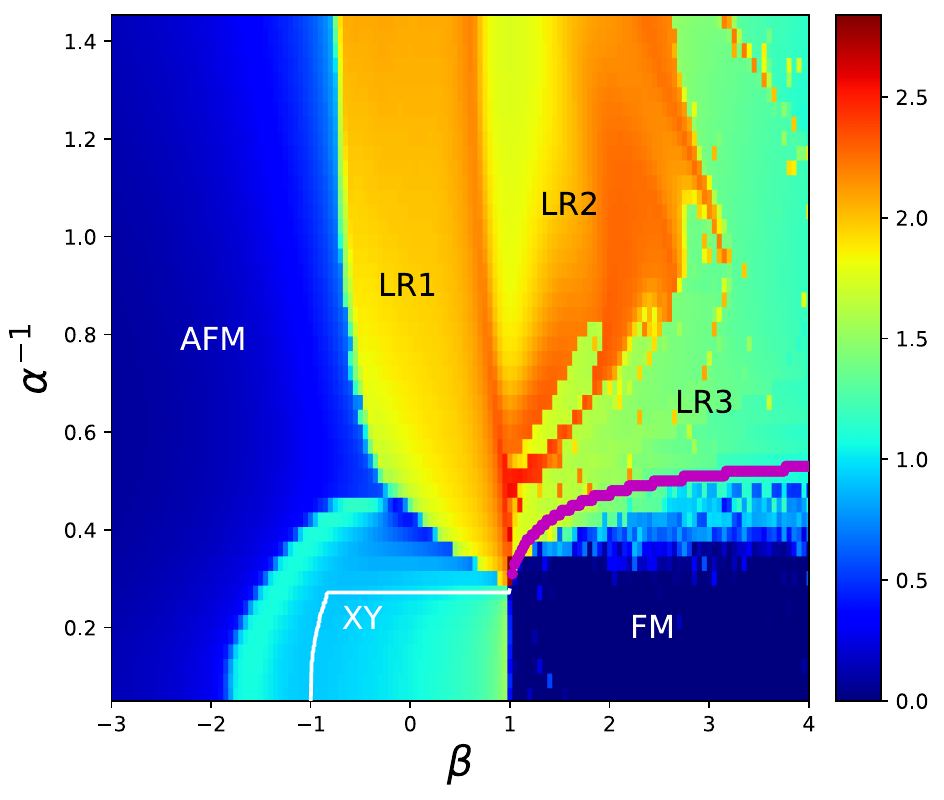}
\caption{\justifying\parfillskip=0pt\relax\label{fig:EE_n100a2a}The quantum phase diagram for Hamiltonian, Eq. \eqref{eq:hamG}, represented by the entanglement entropy \(S_E\), calculated using finite-size DMRG. Anisotropy \(\beta\) varied against LR exponent \(\alpha^{-1}\) for \(L=100\). Six phases are identified and annotated in the parameter sweep. Spin wave analysis identifies the FM-LR3 phase transition, Eq. \eqref{eq:hpfm}, and XY phase boundaries, Eq. \eqref{eq:hpxysolution}, traced in purple and white respectively.}
\end{figure}

\emph{Results}.--- The phase diagram is obtained from DMRG entanglement entropy, Fig.~\ref{fig:EE_n100a2a}, showing that SR dominated phases, FM, XY become unstable at sufficiently strong long-range interactions.  Alongside this boundary, we present the gapless LR1 LR-entangled quantum phase absent from the NN XXZ chain \cite{agrapidis2019} and the non-competing LR XXZ chain \cite{maghrebi2017}. It has \(c=0.997(17)\), a twofold-degenerate ground state manifold and entanglement spectrum, and non-trivial order, identified by non-zero non-local entanglement in the \(S^t\) and \(S^q\), measures of the Quantum Conditional Mutual Information (QCMI). A further phase, LR2, appears at stronger LR coupling, and comparison with the SR and LR limits identifies SR-LR competition as the origin of all three. The LR phases are characterised by central charge, ground state degeneracy, entanglement spectrum, QCMI, and correlation functions. Our findings are summarised in Table~\ref{tab:table}.

\begin{table}[b]
\caption{\label{tab:table}\justifying
Table of results for LR1, LR2, and LR3 that summarise the characteristics of each phase. The degeneracy of certain measures is denoted as \(\mathcal{D}[\;\cdot\;]\).
}
\begin{ruledtabular}
\begin{tabular}{l@{\hspace{1.5em}}rrr}
\textrm{Measure}&
\textrm{LR1}&
\textrm{LR2}&
\textrm{LR3}\\
\colrule
\(S_E\) Scaling & Logarithmic & Plateau & Logarithmic\\
\(c_{est}\) & \(0.997\pm0.017\) & --- & \(1.837\pm 0.020\)\\
\(\Delta _\infty\) & 0 & 0.047 & 0 \\
\(\mathcal{D}[E_0(L, S^z)]\) & 2 & 1 & 2\\
\(\mathcal{D}[E_S]\) & 2 & --- & ---\\
\(S^t\) & \(\sim1.2\) & \(\sim0.8\) &  \(\to0\) \\
\(S^q\) & \(\sim0.1\) & \(\sim0.2\) &  \(\to0\) \\
\(\expval{S^+_i S^-_j}\)& Strong LR & Weak LR & --- \\
\(\expval{S^z_i S^z_j}\)& --- & Strong LR & Non-decaying \\
\end{tabular}
\end{ruledtabular}
 \\
\end{table}

The model is explored with a two-site DMRG algorithm using TeNPy \cite{tenpy} to calculate the low excitation spectrum, with a subspace expansion and a large maximum bond dimension \(\chi_{max}\geq1000\) to ensure convergence. In contrast, models without SR-LR competition can use significantly smaller bond dimensions to accurately capture their ground states, even in the presence of LR correlations \cite{schneider2022}. The phase diagram is calculated using the entanglement entropy \(S_E\), which quantifies the degree of entanglement across a bipartitioned system of equal subsystem sizes and is a good indicator for quantum phase transitions \cite{pollmann2010}. The entanglement entropy is given as \(S_E = -\sum_i \lambda_i^2\log(\lambda_i^2)\), where \(\lambda_i\) are the Schmidt coefficients obtained from the singular value decomposition of the matrix product state representation of the ground state at the central bond. Fig.~\ref{fig:EE_n100a2a} shows the phase diagram parametrised by anisotropy \(\beta\) and interaction strength \(\alpha\).

In the limit at \(\alpha \longrightarrow \infty\), we obtain the SR anisotropic XXZ spin chain, which is exactly solvable by the Bethe ansatz and contains a gapped FM and AFM phase, as well as a gapless XY phase \cite{giamarchi}.  In the LR regime, non-trivial quantum phases LR1 and LR3 can be seen in distinct regions of near uniform \(S_E\). LR2 comprises a region of smoothly varying \(S_E\), but is contained within the visible \(S_E\) discontinuities between phases LR1 and LR3.

Finite-size DMRG calculations show phase boundaries that are smeared \cite{Ueda2008}, seen specifically for the FM phase. Phase boundaries in systems of interacting spins can be explored by utilising a Holstein-Primakoff transformation and examining the dispersion relation of low-lying spin-wave excitations \cite{Holstein1940}.  For the FM phase, we define the vacuum state to be the maximally polarised in the \(+z\) direction. From which, spin-1/2 operators are mapped to bosonic operators by, \(\hat{S}_i^z=\tfrac{1}{2}-\hat{a}_i^\dagger\hat{a}_i \) and \(\hat{S}_i^+ = \hat{a}_i^\dagger (1-\hat{a}_i^\dagger\hat{a}_i)^{\frac{1}{2}}\) \cite{Holstein1940}. In the low-excitation limit \(\expval*{\hat{a}_i^\dagger\hat{a}_i}\ll 1/2\), the transformation is linear, and the Hamiltonian becomes \(\hat{H}_{FM} = \sum_k \omega_k \hat{c}_{k}^{\dagger}\hat{c}_{k}\), with dispersion relation \cite{supmat}
\begin{equation} \label{eq:hpfm}
    \omega_k = 
     \text{Re}(\text{Li}_\alpha (e^{ik})) -2\cos(k) +\beta(2-\zeta(\alpha)), 
\end{equation}
where \(\zeta(\alpha) = \sum_{r=1}^\infty 1 / r^\alpha\) is the Riemann zeta function and \(\text{Li}_\alpha (e^{ik})\) is the poly-log function. For each point in the phase space \((\beta, \alpha^{-1})\), if \(\omega_k>0 \; \forall \; k\in [-\pi, \pi)\), then we consider that point as part of the FM phase. The FM-LR3 boundary corresponds to where LR AFM-interactions destabilise FM order, and is traced in purple, seen in Fig.~\ref{fig:EE_n100a2a}. For the XY phase, a similar Holstein-Primakoff transformation gives a bosonic Hamiltonian with normal and pairing terms \cite{supmat} \(\omega(k)\) and \(\mu(k)\), \begin{align}\label{eq:hpxysolution}
    \omega(k) &= 8-4\zeta(\alpha)+2(1+\beta)(2\cos(k)+\text{Re}(\text{Li}_\alpha(e^{ik}))), \nonumber\\
     \mu(k) &= 2(1-\beta)(\text{Re}(\text{Li}_\alpha (e^{ik})) -2\cos(k)). 
\end{align}
which a Bogoliubov transformation diagonalises to give the energy spectrum \(\varepsilon(k)=(\omega(k)^2-\mu(k)^2)^{\frac{1}{2}}\). The XY ansatz holds while \(\abs{\omega(k)}>\abs{\mu(k)}\) and for sufficiently small \(\alpha^{-1}\), the transition is located by tracking where this inequality fails. The XY phase boundary in the thermodynamic limit is printed in white on Fig.~\ref{fig:EE_n100a2a}.

SR-LR competing orders introduce finite-size effects beyond those of standard LR models \cite{maghrebi2017}, so agreement between DMRG and spin-wave predictions improves with system size and toward \(\alpha\to\infty\) \cite{Gong2016}, as confirmed by recovering the known AFM-XY transition at \(\beta=-1\) \cite{giamarchi}. Spin-wave analysis suggests that the XY-LR1 phase boundary flattens and occurs at constant \(\alpha^{-1}\approx 0.27\), which agrees with the FM-LR3 transformation at \(\beta=1\).

\begin{figure}[h!]
\includegraphics[width=0.45\textwidth]{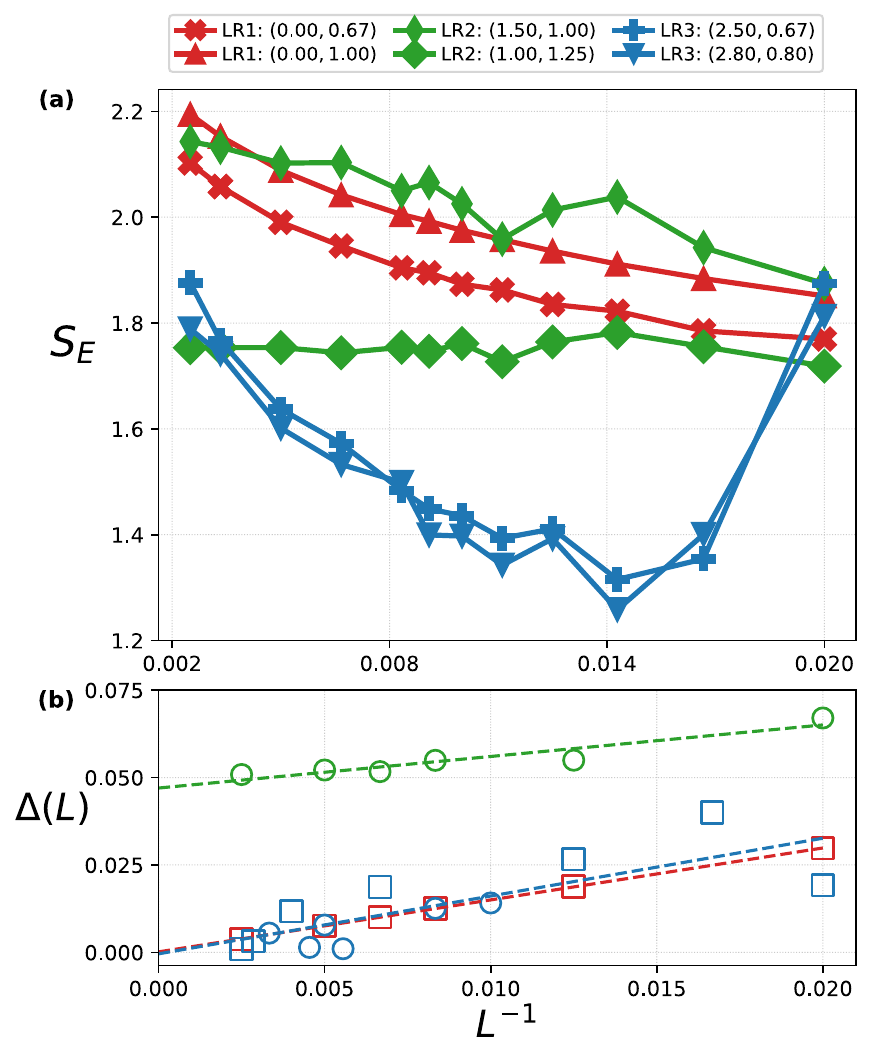}
\caption{\justifying\parfillskip=0pt\relax\label{fig:ee_scaling} (a) Entanglement entropy \(S_E\) scaling with system size \(L^{-1}\) up to \(L=400\) for varying \((\beta, \alpha^{-1})\) parameter values. Logarithmic scaling of \(S_E\) with system size can be seen for the LR1 and LR3 phases, suggesting gaplessness in the thermodynamic limit. (b) The energy gap scaling between the ground state and the first excited state for LR1 (red), LR2 (green), and LR3 (blue). Circle (square) markers indicate the first excited state is found in sector \(\sum_i S_i^z=0 \;(1)\).}
\end{figure}

The LR phases that arise have quantitatively different entanglement properties. The nature of the LR phases is probed by calculating \(S_E\) in the thermodynamic limit, shown in Fig.~\ref{fig:ee_scaling} (a). LR1 and LR3 show logarithmic scaling with system size, indicative of gapless quantum phases. For 1D open-boundary spin chains at criticality, the von Neumann entropy relates to the central charge as \(S(l) = \tfrac{c}{6}\ln[\tfrac{2L}{\pi}\sin\tfrac{\pi l}{L}]+S_0\), for some non-universal constant \(S_0\) \cite{Calabrese2009}. For the half-chain partitioning \(S(L/2)=S_E\), we extract the effective central charge \(c\) from \(S_E = \frac{c}{6}\log(L)\) \cite{Calabrese2009}. We find that for LR1, \(c_{LR1}=0.997(17)\), which aligns with the known conformal field theory (CFT) of the XY phase, \(c_{XY}=1\). Where finite-size effects subside for LR3 (\(L>70\)), the effective central charge is calculated as \(c_{LR3} = 1.837(20)\) does not correspond to a standard CFT value, reflecting the breakdown of conformal invariance by LR interactions \cite{Koffel2012, Vodola2014}.

\(S_E\) scaling identifies LR1 and LR3 as gapless but does not reveal if this coincides with ground-state degeneracy. We resolve this by tracking the ground-state energy manifold across charge sectors \(\sum_i\expval{S_i^z}=\{-L/2, \dots, L/2\}\). The energy gap \(\Delta(L)\) between the ground state manifold and excited states in the thermodynamic limit, \(\Delta_{\infty}\), is extrapolated using a linear fit, Fig.~\ref{fig:ee_scaling} (b). LR1 possess a 2-fold degenerate ground-state in the \(\sum_i\expval{S_i^z}=0\) sector, with a shrinking energy gap. Smaller chain lengths at \(\beta=0\) show this property when \(\alpha^{-1}\to\infty\) \cite{callison2017}. LR2 possess no ground-state degeneracy and remains gapped in the limit \(L\to\infty\). The axis-intercept for LR2 suggests that a gap remains of \(\Delta_{\infty}\approx0.047\), outside of the regime where the Lieb-Schultz-Mattis theorem constraints apply \cite{LSM, Nomura2015, Zhou2024, Ma2024, Yao2022}. The ground state manifold for LR3 swaps between \(\sum_i\expval{S_i^z}=0, 1\) sectors due to nearly-degenerate (\(\mathcal{O}(10^{-5})\)) ground states, where both sectors compete but converge to zero. 

\begin{figure}[h!]
\includegraphics[width=0.41\textwidth]{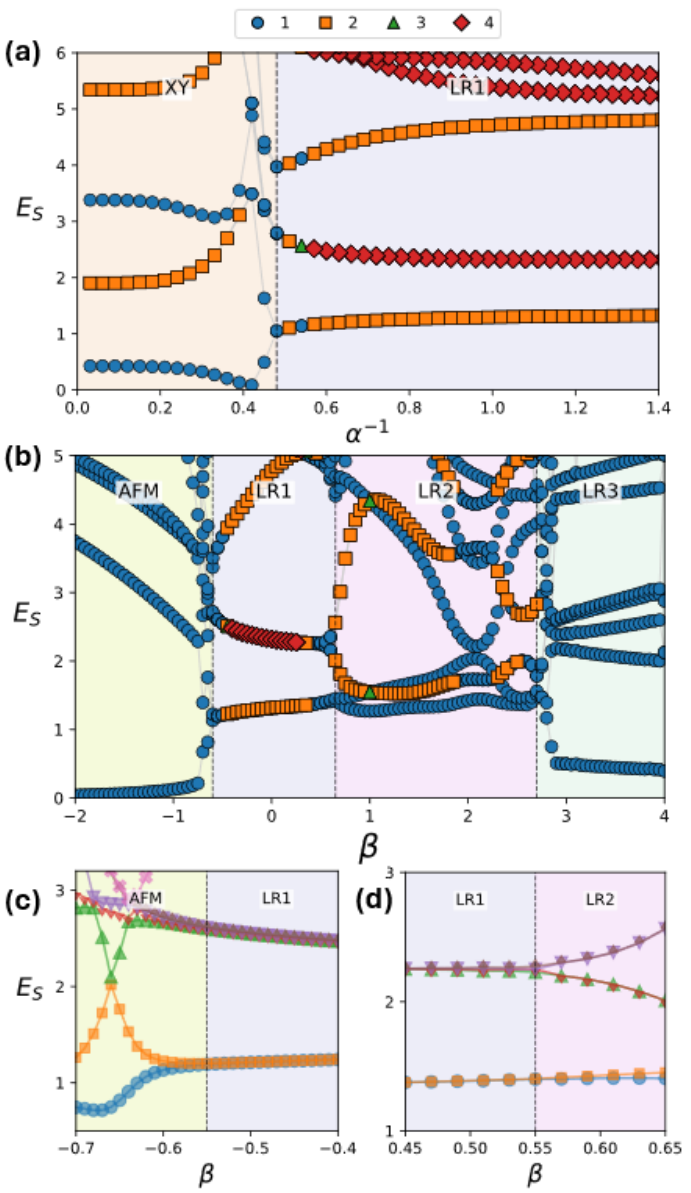}
\caption{\justifying\parfillskip=0pt\relax\label{fig:entspec}Low-lying entanglement spectrum \(E_S\) for (a) \(\beta=0\) showing the XY-LR1 transition, (b) \(\alpha^{-1}=1.2\) showing the AFM-LR1-LR2-LR3 transitions. The legend markers correspond to the degeneracies in (a)-(b). The AFM-LR1 transition (c) and LR1-LR2 transition (d) show each level in \(E_S\) uniquely, highlighting degeneracy formation in LR1.}
\end{figure}

Attributing the two-fold degeneracy of the LR1 ground state to accidental or structural degeneracy can be probed via the entanglement spectrum \(E_S=\{\xi_i\}\), reveals more about the nature of a quantum phase than the entanglement entropy \cite{li2008, pollmann2010, agrapidis2019, schneider2022} and is defined by the Schmidt values \(\{\lambda_i\}\) at the central bond by the relation \(\lambda_i = \exp(-\xi_i/2)\) \cite{li2008}. The parameter space is swept to reveal the quantum phase transitions and degeneracies in Fig.~\ref{fig:entspec}. The XY-LR1 transition at \(\beta=0\), Fig.~\ref{fig:entspec} (a), presents as a collapsing of \(\{\xi_i\}\) into a degenerate spectrum. This characterises LR1, as the LR-regime parameter sweep at \(\alpha^{-1}=1.2\), Fig.~\ref{fig:entspec} (b), reveals the phase transition across AFM-LR1-LR2-LR3. While LR2 and LR3 possess some accidental degeneracies, LR1 presents with a complete minimum twofold-degeneracy in the bulk, indicative of a non-trivially ordered quantum phase \cite{Fidkowski2010, li2008, pollmann2010}. Increasing the strength of the LR interactions in the uniform-sign FM-interacting LR XXZ spin chain does not result in degenerate entanglement spectra in the continuous-symmetry-breaking phase \cite{schneider2022}. Since LR1 appears only when FM-interacting NN terms compete with the AFM-interaction tail, we attribute the degeneracy to this competition \cite{agrapidis2019, supmat}. The degeneracy persists under a shifted entanglement cut \cite{supmat}.

\begin{figure}
  \centering
  \begin{tikzpicture}
    \node[anchor=south west, inner sep=0] (img)
      {\includegraphics[width=0.45\textwidth]{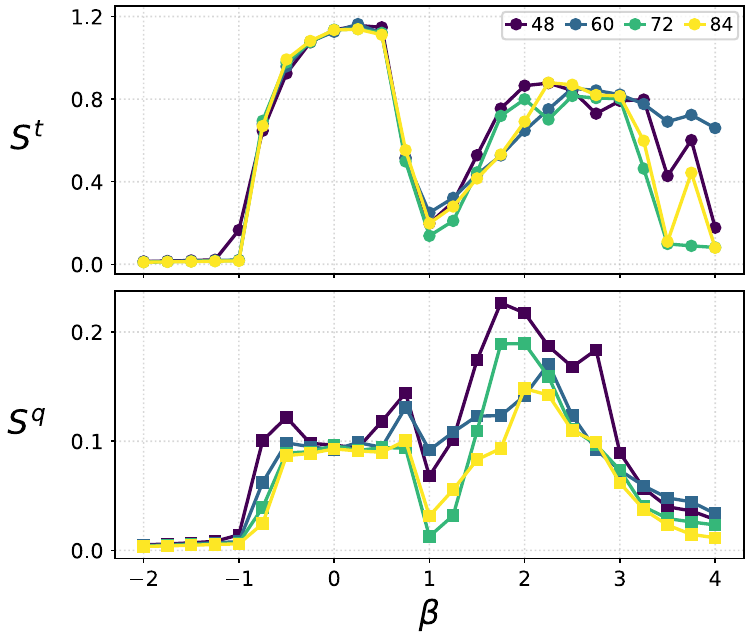}};
    \begin{scope}[x={(img.south east)}, y={(img.north west)}]
      \node[anchor=north west, inner sep=2pt] at (0.16, 0.985) {\textbf{(a)}};
      \node[anchor=north west, inner sep=2pt] at (0.16, 0.540) {\textbf{(b)}};
    \end{scope}
  \end{tikzpicture}
  \caption{\justifying\parfillskip=0pt\relax\label{fig:qcmi}QCMI for both \(S^t\) (a) and \(S^q\) (b) partitioning schemes for \(\beta\) sweep with fixed \(\alpha^{-1}=1.2\).}
\end{figure}

The QCMI is introduced as a non-local entanglement probe of the ground state to identify non-trivial quantum phases \cite{qiqm, Maiellaro2022, Maiellaro2023}, and non-trivial phase transitions even in small system sizes \cite{Chen2015}. The QCMI \(I(A\!:\!C|B)\) is calculated, \begin{equation}
    I(A\!:\!C|B)=S(AB)+S(BC)-S(B)-S(ABC),
\end{equation}
with entanglement entropies \(S(l)\) of the subsystems \(l\in\{AB, BC, B, ABC\}\), where \(I(A\!:\!C|B)\) is defined on a tripartite state \(S^t\) and quadripartite state \(S^q\) of subsystem sizes \(L/3\) and \(L/4\) respectively \cite{Chen2015, Zeng2016, Zeng2015}. \(S^t\) and \(S^q\) are calculated up to \(L=84\) in the LR-regime \(\alpha^{-1}=1.2\) and presented in Fig.~\ref{fig:qcmi}.

\begin{figure}
\includegraphics[width=0.45\textwidth]{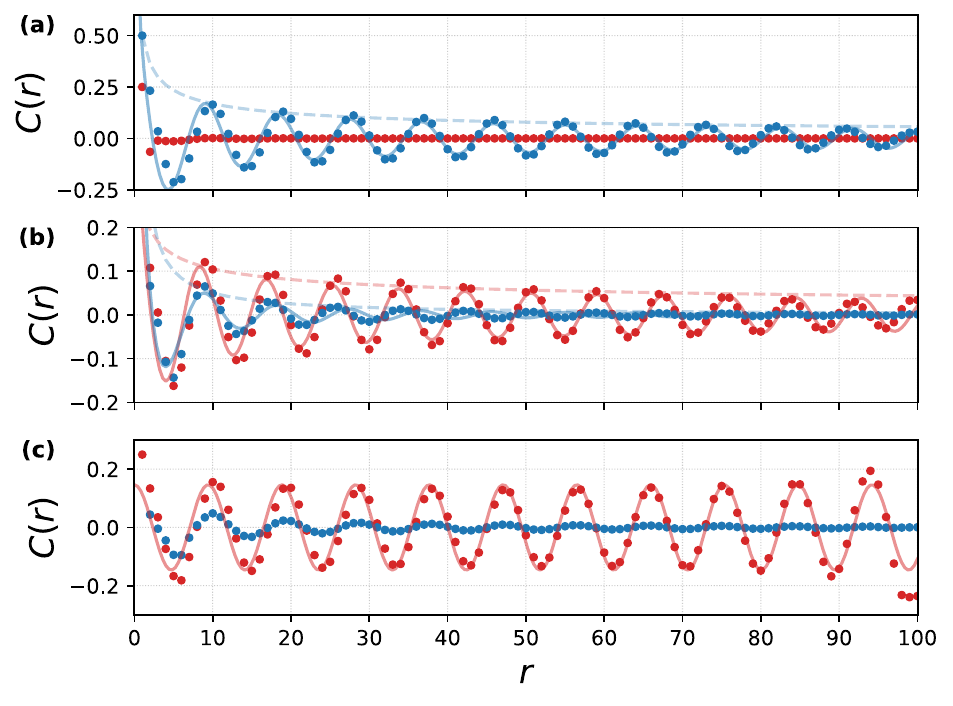}
\caption{\justifying\parfillskip=0pt\relax\label{fig:oscils} Spin-spin correlations \(\expval*{S^z_i S^z_j}\) (red) and \(\expval*{S^+_iS^-_j}\) (blue) across the ground state for \(L=100\) in the (a) LR1, (b) LR2, and (c) LR3 quantum phases. \((\beta, \alpha^{-1})\) coordinates are (0.15, 0.81), (1.9, 0.93), (2.5, 0.6) respectively. Enveloping power-law decay functions are fitted in dashed lines, and oscillations are traced in solid lines.}
\end{figure}

The AFM-LR1 phase transition can be seen at \(\beta=-1\), where \(S^t\) saturates at \(\sim 1.2\), whilst the LR1-LR2 phase transition is identified as the sharp decrease in \(S^t\). Non-zero \(S^q\) for \(-1<\beta<3\) indicates that the ground states of LR1 and LR2 possess LRE \cite{Zheng2026}. Dominant \(S^t\) in these phases indicates the presence of symmetry-breaking order \cite{qiqm}. In LR1, \(S^t \approx 1.2\) and \(S^q \approx 0.1\) are finite. In the adjacent XY phase, \(S^t\) is an order of magnitude smaller \cite{supmat}. LR3, \(\beta>3\), shows monotonically decreasing non-local entanglement. The spin-spin correlation functions of each phase reveals the physical manifestation of the entanglement presented by the QCMI. The correlation functions \(\expval*{S^z_i S^z_j}\) and \(\expval*{S^+_i S^-_j}\) as a function of \(r=\abs{i-j}\) are shown in Fig.~\ref{fig:oscils}. Oscillations in each spin-spin correlator identify the phases directly. LR1, Fig.~\ref{fig:oscils} (a), shows oscillating \(\expval*{S^+_i S^-_j}\) correlations enveloped by a weakly-decaying power-law. Power-law decay signals the XY phase, but oscillations are unique to LR1. LR2, Fig.~\ref{fig:oscils} (b), shows weaker \(\expval*{S^+_i S^-_j}\) but stronger \(\expval*{S^z_i S^z_j}\) correlations. LR3, Fig.~\ref{fig:oscils} (c), shows no decay in oscillation strength, as large \(\beta\) and sufficient \(\alpha^{-1}\) minimise LR AFM costs by favouring SR quasi-FM alignment, broadening the LR3-FM boundary. At \(\beta\gg 1\) this recovers the ANNNI model with no transverse field \cite{Cea2024, Fumani2021}, which reveals that LR3 is an XX extension to the antiphase \cite{Torre2025}, with an additional LR tail.

\emph{Conclusions}.--- We have studied a spin-1/2 XXZ chain with competing FM NN and AFM power-law interactions, and found that this competition generates three phases absent from the NN chain, identified using information-theoretic measures such as the entanglement spectrum and QCMI. Spin-wave theory locates the destabilisation boundary of FM order to LR AFM order, and DMRG confirms the prediction. Previously established phases are modified in the LR limit \(\alpha^{-1}\gtrsim 0.27\). LR1 and LR2 both host LRE. For \(-1<\beta<1\), LR1 is a gapless symmetry-breaking phase with twofold-degenerate ground states, generated from the unique interplay between SR and LR interactions. It shows that the phase of Ref. \cite{callison2017}, found for all-to-all coupling on small chains, persists in the thermodynamic limit and is robust to \(S^zS^z\) perturbations. The twofold degeneracy of its entanglement spectrum distinguishes it from the continuous symmetry-breaking phases of the uniform-sign LR XXZ model \cite{maghrebi2017, schneider2022}. LR2 is a strongly correlated, LR-entangled phase that evades the LSM theorem. LR3 is an XX-type counterpart of the ANNNI antiphase. These results show that competition between SR FM and LR AFM interactions can produce a gapless LRE phase with no counterpart in either limit. The oscillating correlations and twofold-degenerate entanglement spectrum of LR1 offer signatures for platforms with tunable-sign power-law couplings, such as Rydberg arrays.

E. Gi. acknowledges support from British Council (ISPF) grant number 47650978 and from the Engineering and Physical Sciences Research Council grant number EP/Z000505/1. E. Gr. acknowledges support from the Israeli Ministry of Innovation, Science and Technology under the UK/Israel Collaboration Grant QuantumGate.

\bibliography{mybib}

\end{document}